\documentclass[
 aip,
 amsmath,amssymb,
preprint,
]{revtex4-1}

\usepackage{graphicx}
\usepackage{dcolumn}
\usepackage{bm}

\usepackage[utf8]{inputenc}
\usepackage[T1]{fontenc}
\usepackage{mathptmx}
\usepackage{etoolbox}
\usepackage{enumitem}

\usepackage{color}

\makeatletter
\def\@email#1#2{
 \endgroup
 \patchcmd{\titleblock@produce}
  {\frontmatter@RRAPformat}
  {\frontmatter@RRAPformat{\produce@RRAP{*#1\href{mailto:#2}{#2}}}\frontmatter@RRAPformat}
  {}{}
}
\makeatother
\begin{document}

\title[Physics-informed Bayesian inference of external potentials in classical density-functional theory]{Physics-informed Bayesian inference of external potentials in classical density-functional theory}
\author{Antonio Malpica-Morales}
\email{a.malpica-morales21@imperial.ac.uk}
\affiliation{
Department of Chemical Engineering, Imperial College, London SW7 2AZ, United Kingdom
}%
\author{Peter Yatsyshin}
\email{pyatsyshin@turing.ac.uk}
\affiliation{
Department of Chemical Engineering, Imperial College, London SW7 2AZ, United Kingdom
}%
\affiliation{
The Alan Turing Institute, London NW1 2DB, United Kingdom
}%
\author{Miguel A. Dur\'an-Olivencia}
\email{miguel@vortico.tech}
\affiliation{
Department of Chemical Engineering, Imperial College, London SW7 2AZ, United Kingdom
}%
\affiliation{
Research, Vortico Tech, M\'alaga 29100, Spain
}%
\author{Serafim Kalliadasis}
\email{s.kalliadasis@imperial.ac.uk}
\affiliation{
Department of Chemical Engineering, Imperial College,  London SW7 2AZ, United Kingdom
}%

\date{\today}

\begin{abstract}
\textbf{ABSTRACT}
\\
\\
The swift progression and expansion of machine learning (ML) have not gone 
unnoticed within the realm of statistical mechanics. 
In particular, ML techniques have attracted attention by the classical 
density-functional theory (DFT) community, as they enable automatic discovery
of free-energy functionals to determine the equilibrium-density profile of a 
many-particle system. 
Within classical DFT, the external potential accounts for the interaction of the 
many-particle system with an external field, thus, affecting the density 
distribution. 
In this context, we introduce a statistical-learning framework to infer the 
external potential exerted on a classical many-particle system. 
We combine a Bayesian inference approach with the classical DFT apparatus to 
reconstruct the external potential, yielding a probabilistic description of the 
external potential functional form with inherent uncertainty quantification. 
Our framework is exemplified with a grand-canonical one-dimensional
classical particle ensemble with excluded volume interactions in a confined 
geometry. 
The required training dataset is generated using a Monte Carlo (MC) simulation 
where the external potential is applied to the grand-canonical ensemble. 
The resulting particle coordinates from the MC simulation are fed into the 
learning framework to uncover the external potential. 
This eventually allows us to characterize the equilibrium density profile of the 
system by using the tools of DFT. 
Our approach benchmarks the inferred density against the exact one calculated
through the DFT formulation with the true external potential. 
The proposed Bayesian procedure accurately infers the external potential and
the density profile. 
We also highlight the external-potential uncertainty quantification conditioned 
on the amount of available simulated data.
The seemingly simple case study introduced in this work might serve as a 
prototype for studying a wide variety of applications, including
adsorption, wetting, and capillarity, to name a few.
\end{abstract}

\maketitle

\section{Introduction}

The advancement of statistical learning techniques, coupled
with more affordable computing power and access to large datasets,
has fueled the recent rapid growth of machine learning (ML).~\cite{Marsland2014}
ML has achieved unprecedented progress in a large spectrum of
fields including image recognition, natural language processing,
and biological-structure prediction. 
At present, incorporating ML into science and engineering is seen 
as essential for advancing and expediting research.

In the field of materials science, in particular, an interdisciplinary
subject spanning the physics, chemistry and engineering
of matter, and industrial manufacturing processes, driven by
a fast and rational approach to design new materials, ML has
become an essential toolkit in computational materials modeling
and molecular simulations.~\cite{Schmidt2019, Schleder2019, Moosavi2020}
In this field, ML facilitates the discovery of novel molecules, materials, or 
systems exhibiting desired characteristics.
In particular, a very important contribution to the
field is the accurate representation of potential-energy surfaces
for atomistic computer simulations.
This problem has attracted
considerable attention with recent developments of the so-called
ML potentials.~\cite{Bartok2013, Behler2016, Tong2020}
These potentials provide a direct functional
relation between the atomic positions and the associated energy
configuration by fitting a training set of electronic structure calculations.
However, ML potentials utilize elements from quantum
density-functional theory (DFT), and therefore their applicability is
necessarily restricted to electron-size scale.

Quantum DFT is built upon a theorem formulated by Mermin.~\cite{Mermin1965}
It establishes that the grand potential of an electron gas subjected to a
one-body potential can be expressed as a distinctive functional of the
local density.
This theorem is a finite-temperature extension of the
Hohenberg–Kohn theorem~\cite{Hohenberg_1964} that operates at zero-temperature conditions.
These two theorems provide the foundations for the DFT in
\textit{ab initio} quantum mechanical calculations.

Classical DFT, which deals with molecular-scale interactions
in many-particle systems, such as liquids and colloids, has its
roots on the same fundamental theorems.~\cite{Lutsko2010}
It is a statistical-mechanical
framework in which the free energy of a many-particle
system is written as a functional of the particle density, $\rho(\mathbf{r})$.
The system’s equilibrium is then determined by the extrema of
the free-energy functional, which represent the density profiles at
equilibrium.~\cite{Evans1979}
The density $\rho(\mathbf{r})$ itself can be understood as the
probability density function describing the likelihood of a particle
being located in the proximity of the position-vector $\mathbf{r}$.
By means of this one-particle representation of the density, properties of 
matter, such as electric charge, magnetization, or pressure, are derived
by weighing the microscopic interactions between the constituent
particles.

While classical DFT is a popular statistical-mechanical framework,
it hinges on the exact formulation of the Helmholtz free
energy, which remains undetermined for certain system configurations.
This barrier can be overcome by simplifying the intermolecular
interactions through coarse-graining techniques yielding an
approximate free-energy functional. 
Such approximations have been widely explored in several realistic many-particle 
systems, accounting for effects, such as geometry, phase transitions, 
nucleation, and multiple components.\cite{Yatsyshin2013,
Yatsyshin2017, Lutsko2019}

Previously, we provided an example of an ML application
to quantum DFT. 
The applicability of ML to classical DFT has
primarily focused on the discovery of true free-energy functionals
as an alternative to the human-designed functionals.~\cite{Pederson2022}
Ref.~\onlinecite{Cats2021} combines the classical DFT formalism with ML, 
specifically kernel-density regression, to approximate free-energy functionals.
In that work, a collection of grand-canonical Monte Carlo
(MC) simulations at different chemical and external potentials
in a planar geometry is used as a training set to obtain free-energy
functionals for a Lennard-Jones fluid at supercritical temperature.
The learning process involves measurement of the error
between the predicted density profiles coming from the DFT apparatus
in which the ML free-energy functional is embedded and
the empirical results of the MC simulations, which are taken
as the ground truth. 
Despite the accuracy obtained in the free-energy
functional reconstruction and the ability to extract additional
physics-related parameters such as thermodynamic bulk
quantities and two-body correlations, the authors of Ref.~\onlinecite{Cats2021} 
highlight some spurious results attributed to overfitting in the training
process.

The authors of Refs.~\onlinecite{Lin2019} and \onlinecite{Lin2020} 
were the first to employ 
neural networks for free-energy functional reconstruction; particularly
the free-energy functional for hard rods (HR) and a Lennard-Jones
fluid in one dimension (while a comprehensive perspective of ML
methods in DFT, including neural networks, is given in 
Ref.~\onlinecite{delasHeras2023}).
Neural networks are very good at handling complex inputs, such
as images,~\cite{Carillo2021} and perform as universal approximators
\cite{Hornik1989} yielding
high accuracy in function reconstruction. 
However, they effectively operate as a black-box model being difficult to 
interpret and requiring large datasets to approximate high-dimensional 
functions. 
In addition, neural networks do not generate uncertainty quantification
by default. 
They produce a single length-fixed array instead of a collection of possible 
outputs weighted by probabilities as in the case of Bayesian inference. 
There exist techniques to incorporate uncertainty quantification into neural networks, such as MC dropout \cite{Gal2016} and ensemble method. \cite{Lakshminarayanan2017}
However, both methods pose some challenges. 
MC dropout relies on the empirical definition of 
the neural network architecture, making the uncertainty estimation
potentially sensitive to different network configurations. 
On the other hand, the ensemble method might be computationally
demanding as it requires training the same neural network model
a predefined number of times. 
In addition, the choice of this predefined number introduces another 
hyperparameter that affects the uncertainty quantification process.

In contrast to neural networks and traditional ML techniques,
non-parametric Bayesian inference from canonical and grand-canonical
simulations provides uncertainty quantification of the
predictions for free-energy functionals, as demonstrated in
Ref.~\cite{Yatsyshin2022}. 
(The same study also implemented adjoint techniques, allowing the
use of gradient-based samplers, such as Hamiltonian MC). 
Following the Bayesian approach also, Ref.~\onlinecite{Yousefzadi2020}
utilized simulated data from molecular dynamics to approximate the effective 
potential governing the random walk of an atom in a lattice.

The employment of classical DFT to obtain $\rho(\mathbf{r})$ might be a
challenging task if system particles interact through fine-detailed
potentials or if there exists an external potential acting on the
many-particle system. 
External fields might modify the mechanical, electrical, or optical properties 
of systems in equilibrium with a bulk fluid having profound impacts on fields 
such as separation processes, crystallization, or nanofluidics, to name a few. 
Therefore, the external potential stands as one of the central variables to
accommodate classical DFT,~\cite{Lowen2002} given that it might alter the 
equilibrium density $\rho(\mathbf{r})$.
It is noteworthy that the external-potential 
effects on the adsorption and phase transitions occurring at the
substrate–fluid interface have been researched by \textit{ab initio} analytical 
calculations (including DFT),~\cite{Yatsyshin2017, Yatsyshin2018} molecular 
simulations,~\cite{Snook1978, Brunet2009, Brunet2010} or 
both.~\cite{Henderson1984, Segura2001}

Rather than assessing the external-potential effects on properties
of matter, our aim is to uncover the external potential acting
on a many-particle system by means of simulated data. 
The data corresponds to system-configurations observations that on average
resemble the equilibrium-density profile of the many-particle system
under the influence of the external potential. 
As a prototypical example, we consider a many-particle system with excluded 
volume interactions, which are ubiquitous in many applications, where
a predefined external potential is exerted. The advantage of our
prototype is that the associated free-energy functional is known analytically.
Thus, we can leverage the classical DFT apparatus to statistically
learn the external potential in a physics-informed ML formulation.
Regarding the ML technique, we adopt a Bayesian inference
framework, precisely because of its ability to provide full uncertainty
quantification, as already alluded to. 
As we will see, uncertainty quantification depends solely on the amount of 
available data. 
It is based on seamless uncertainty propagation through all modeling
hierarchy levels, thereby facilitating a judicious interpretation of the
results.

\section{External potential on the grand-canonical Ensemble}

In statistical mechanics, the one-dimensional (1D) functional
of HR has a tractable analytical expression.~\cite{Evans1992}
The same expression can be obtained using particle simulations over a 
computational ensemble configured with identical parameters to the theoretical 
setup.
This equivalence is also valid when an external field or potential
is exerted over the many-particles system. 
The objective of this work is to learn the external potential applied over a 1D 
purely repulsive HR system by using as empirical data the particles’ coordinates
obtained from molecular simulations. 
For the sake of simplicity, we restrict our attention to a 1D HR system confined 
between two walls.
The configuration resembles a fluid inside a pore. 
In this context, we calculate the exact equilibrium density resulting from DFT. %
The obtained density is used as the ground truth in the external-potential
learning process. 
We then present the MC simulation that generates the particle configurations that are used as the learning dataset.

\subsection{Exact density profile from DFT}

First, we define an external potential $V(x)$ and solve the direct
problem of statistical mechanics, i.e., finding the probability density $\rho(x)$ by minimizing the grand potential energy density functional,
\begin{align}
    \Omega[\rho] \equiv \mathcal{F}[\rho] + \int dx \rho(x) \left( V(x) - \mu \right),
    \label{eq:grand-potential}
\end{align}
where $\mathcal{F}$ is the free-energy functional and $\mu$ is the chemical potential of the particle reservoir.
The minimum on $\Omega$ can be found through the functional derivative of $\Omega$,
\begin{align}
    \frac{\delta\Omega[\rho]}{\delta\rho(x)} \equiv \frac{\delta\mathcal{F}[\rho]}{\delta\rho(x)} + V(x) - \mu = 0,
    \label{eq:euler-lagrange}
\end{align}
by taking the functional derivative of the free energy.~\cite{Tarazona2008}

\begin{figure*}
\includegraphics{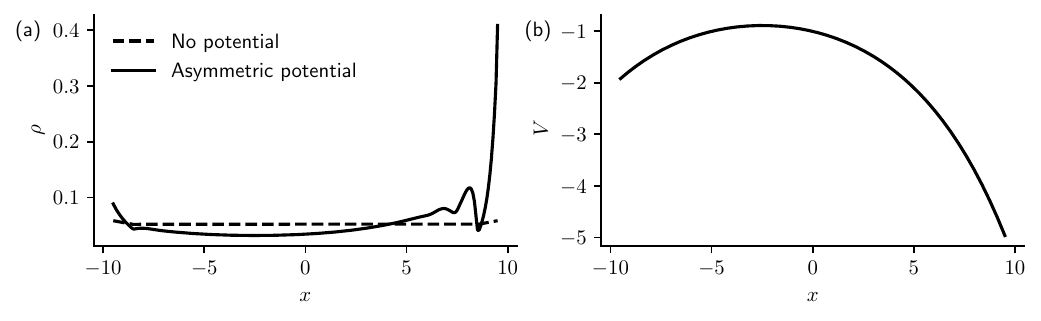}
\caption{\label{fig:exact_dft_potential}(a) Density profiles for the grand-canonical 1D HR ensemble with and without external potential. The ensemble is defined with pore length $L=20$, chemical potential $\mu = -2$ and HR width $2R=1$. (b) External asymmetric potential implemented on the grand-canonical 1D HR ensemble as in Eq.~\eqref{eq:true-potential} with $\epsilon=2, r=5$.}
\end{figure*}

In what follows, all quantities will be reported in reduced units.
We consider a grand-canonical 1D HR ensemble embedded in a
pore of length $L=20$ with a chemical potential $\mu=-2$ and HR width $2R=1$.
Figure~\ref{fig:exact_dft_potential}(a) superimposes the exact $\rho(x)$ 
obtained in two different scenarios for our ensemble configuration: 
One scenario assumes the absence of an external potential, an ideal case, and the
other one considers an asymmetric potential within the pore.
The employed asymmetric potential is shown in Figure~\ref{fig:exact_dft_potential}(b).
This potential is defined as
\begin{align}
    V(x) = -\epsilon\left[ \exp((x - L/4)/r) + \exp((-x - L/2)/r) \right],
    \label{eq:true-potential}
\end{align}
with $\epsilon = 2,$  and $r=5$.

The value $\mu=-2$ defined in the absence of external potential
makes the system very diffuse, as the average number of interacting
particles is $\langle N \rangle = 2.05$.
This small number of particles, taking
also into account the pore dimension, yields a low density profile.
In fact, the layering effect close to the side walls is insignificant and
the system behaves like bulk fluid along the whole pore. 
In contrast, the external potential increases the average number of particles up 
$\langle N \rangle = 6.32$, strengthening the effects of the side walls, 
especially for the right wall at $x=10$, where the potential intensity is larger
compared to the left wall at $x=-10$.

\subsection{MC simulation}

The particle coordinates $\{ x_i\}_{i=1}^{M}$ used to learn the external
potential are simulated using the grand-canonical MC 
algorithm.~\cite{Frenkel2001}
This method allows us to simulate particle coordinates $x$ of HR of radius $R$ 
along the line $L$ considering an external potential $V(x)$ at a given $\mu$.
This technique is widely used in adsorption studies,
so-called the $(\mu,\upsilon,T)$ ensemble in the literature, given that we are 
imposing $\mu$ over a predefined geometry with system volume $\upsilon$ (for 
the 1D system in our case, the characteristic scale is $L$) at a given temperature $T$.
In the simulations, we use reduced units for $T$, obtaining $T = 1$ and $\beta = 1/T = 1$.
The grand-canonical MC algorithm is a computationally efficient and accurate 
method for our purposes, given the straightforward nature of our prototypical
system. 
The simulation steps are as follows:

\begin{enumerate}
    \item Draw a valid random configuration of particle coordinates $\mathcal{X}_1 = (x_1, \ldots, x_{N_1})$ for $1 \leq N_1 \leq L/2R$. Run step \ref{it:attempt} until a desired number of configurations $N_{\text{conf}}$ is obtained.
    \item\label{it:attempt} Attempt to displace a particle (step 3) or exchange a particle (step 4) with the reservoir maintaining a ratio of $npav/nexc$, where $npav$ and $nexc$ are the average numbers of attempts to displace 
    particles and exchange particles, respectively, per cycle.
    \item\label{it:displace} Displace a particle from $\mathcal{X}_j = (x_1, \ldots, x_{N_j})$:
    \begin{enumerate}[label*=\arabic*.]
        \item Choose a particle randomly $x_i, 1 \leq i \leq N_j$.
        \item Compute the potential of current configuration $V(x_i)$.
        \item Give the particle a random displacement, $x^{'}_i = x_i + 2 \times (rand - 0.5)$, where $rand$ is a sample drawn from a uniform distribution $U(0, 1)$.
        \item Compute the potential of new configuration $V(x^{'}_i)$.
        \item Accept the new configuration $\mathcal{X}_{j+1} = (x_1, \ldots, \,  x^{'}_i, \ldots, \, x_{N_j} )$ with probability:
        \begin{align}
        P(\mathcal{X}_j \rightarrow \mathcal{X}_{j+1}) = \min(1, \exp(-\beta[V(x^{'}_i) - V(x_i)]).
        \label{eq:displacement-probability}
        \end{align}
        If $\mathcal{X}_{j+1}$ is rejected, $\mathcal{X}_{j}$ is retained.
        \item Return to step 2.
    \end{enumerate}

    \item Exchange a particle with the reservoir: Choose whether inserting or removing a particle with equal probability. If the outcome is to insert a particle go to step \ref{it:insert}, to remove a particle go to step \ref{it:remove}
        \begin{enumerate}[label*=\arabic*.]
        \item\label{it:insert} Insert a particle into $\mathcal{X}_j = (x_1, \ldots, x_{N_j})$:
            \begin{enumerate}[label*=\arabic*.]
            \item Add a new particle at random position, $x_{N_j +1}$.
            \item Compute the potential of the new particle $V(x_{N_j +1})$.
            \item Accept the new configuration $\mathcal{X}_{j+1} = (x_1, \ldots, x_{N_j}, x_{N_j + 1})$ with probability:
            \begin{align}
            P(\mathcal{X}_j \rightarrow \mathcal{X}_{j+1}) = \min\left(1, \frac{L}{N_j + 1} \exp( \beta [ \mu - V(x_{N_j +1}) ])\right).
            \label{eq:insert-particle-probability}
            \end{align}
            If $\mathcal{X}_{j+1}$ is rejected, $\mathcal{X}_{j}$ is retained.
            \item Return to step 2.
            \end{enumerate}
        \item\label{it:remove} Remove a particle from $\mathcal{X}_j = (x_1, \ldots, x_{N_j})$:
            \begin{enumerate}[label*=\arabic*.]
                \item Choose a particle randomly $x_i, 1 \leq i \leq N_j$.
                \item Compute the potential of the selected particle $V(x_i)$.
                \item Accept the new configuration $\mathcal{X}_{j+1} = (x_1, \ldots, x_{N_j}) - \{ x_i \}$ with probability:
                \begin{align}
                P(\mathcal{X}_j \rightarrow \mathcal{X}_{j+1}) = \min\left(1, \frac{N_j}{L} \exp( -\beta [ \mu - V(x_{i}) ])\right).
                \label{eq:remove-particle-probability}
                \end{align}
                If $\mathcal{X}_{j+1}$ is rejected, $\mathcal{X}_{j}$ is retained.
                \item Return to step 2.
            \end{enumerate}
        \end{enumerate}
\end{enumerate}

In the above steps a number of constraints are satisfied. 
Because the HR behave as purely repulsive particles, if during a particle
displacement (step \ref{it:displace}) or particle insertion (step \ref{it:insert}), there is a volume overlap between the HR, the new 
configuration is automatically rejected going back to step 2.
In addition, when there are no particles during the simulation steps, i.e., $N_j = 0$, the particle displacement (step \ref{it:displace}) or the particle deletion (step \ref{it:remove}) is automatically skipped.

\begin{figure}
\includegraphics[width=0.5\textwidth]{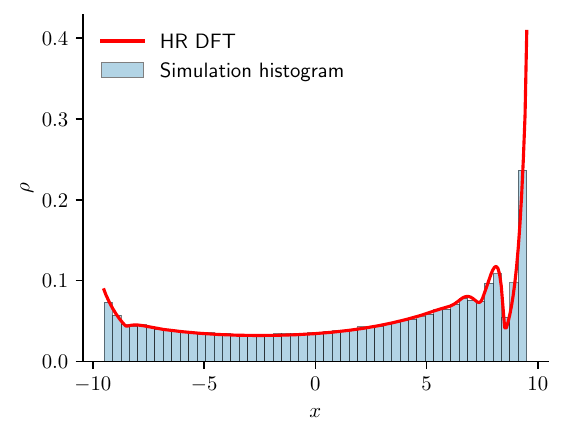}
\caption{\label{fig:exact_dft_simulation} Comparison between the density profile obtained from the exact analytical HR DFT and the normalized histogram obtained from the grand-canonical MC simulation. The simulation runs for $N_{\text{conf}} = 3 \times 10^5$ giving $M \approx 1.8 \times 10^6$ particle configurations. The ensemble parameters for both techniques is $L=20, 2R=1, \mu=-2$ considering the external potential $V(x)$ as in Eq.~\eqref{eq:true-potential} with $\epsilon=2, r=5$.}
\end{figure}

The algorithm runs an undefined number of iterations until
obtaining the desired dataset $\mathcal{D} = (\mathcal{X}_1, \mathcal{X}_2, \ldots, \mathcal{X}_{N_{\text{conf}}} )$, which gives us the particle coordinates $\{ x_i \}^{M}_i$.
When $M$ is large enough, the normalized histogram of $\{ x_i \}^{M}_i$ and the exact solution $\rho(x)$ from Eq.~\eqref{eq:euler-lagrange} converge as shown in Fig.~\ref{fig:exact_dft_simulation}.

\section{Bayesian Inference of the external potential}

Up to this point, we have computed the density profile $\rho(x)$ of
our grand-canonical 1D HR ensemble with external potential using
the DFT formalism and the MC simulation.
With both techniques, $\rho(x)$ is obtained having a perfect knowledge of the 
external potential applied, $V(x)$.
In what follows, we aim at treating the external potential
as an unknown function to be inferred. 
The exact $\rho(x)$ obtained from Eq.~\eqref{eq:euler-lagrange}, with the 
external potential $V(x)$ applied, represents the
target density profile for our external-potential learning problem.
At this stage, we deal with the inverse problem of statistical mechanics. \cite{Chayes1984}
In this inverse problem, we use a set of particle coordinate
observations to approximate the unidentified $V(x)$.
The approximated $V(x)$ is then inserted in the DFT formulation to recover $\rho(x)$.
The greater the precision in determining $V(x)$, the more reliable representation of $\rho(x)$ will be.

The external-potential learning process involves using partial
information coming from the MC simulated particle configurations.
Our statistical learning framework is based on Bayesian inference.
It allows us to approximate a probability distribution over
possible external-potential functions conditioned by the supplied
simulated data. 
We first outline the Bayesian inference methodology
to obtain a probabilistic representation of the unknown
external potential. 
We then comment on the algorithm used to 
construct the probabilistic formulation and discuss the results
obtained.

\subsection{External potential learning procedure}

After simulating the grand-canonical 1D HR ensemble and
obtaining the exact $\rho(x)$ from the classical DFT approach, our goal is
to learn the external potential exerted in the ensemble. 
For this purpose, we use the \textit{in silico} experimental data as the 
training dataset to recover the external potential through Bayesian inference, 
keeping the exact DFT-related $\rho(x)$ as the ground truth to benchmark our model.
Bayesian inference is a statistical technique that provides
a probability description over a set of observable data.~\cite{Gelman2004}
Typically, this involves a probability distribution on the parameters of a 
predefined model that supposedly originates from the input data. 
One of the main advantages of Bayesian inference is the straightforward
uncertainty quantification allowing the judgment of statistical conclusions
and model selection.~\cite{Girolami2008}
The following formula summarizes the essence of Bayesian inference:
\begin{align}
    P(Q | \mathcal{D}) \propto P(Q) P(\mathcal{D}|Q), \label{eq:bayesian-inference}
\end{align}
where $P(Q | \mathcal{D})$ is the posterior distribution of the model parameters $Q$ conditioned on the observed data $\mathcal{D}$, $P(Q)$ is the prior distribution assumed on $Q$ and $P(\mathcal{D}|Q)$ is the sampling or likelihood distribution.
Therefore, Bayesian inference is a suitable framework
to propagate the uncertainty of our prior assumptions in the quest
for the unknown exerted potential on the many-particle ensemble.

To set up our Bayesian framework to uncover the external
potential, we first construct an analytical representation of the
unknown potential. 
We select a mixture model of Gaussian radial basis functions (RBF) as an 
adequate external potential, capable of generating smooth functions,
\cite{Fornberg2011}
\begin{align}
\overline{V}(x) = \sum\limits_{i=1}^{c} Q^{1}_{i}\exp(-(x-z_i)^2/\exp(Q^{2}_{i})), \label{eq:rbf-potential}
\end{align}
where $Q^{1}_{i}, Q^{2}_{i} \in \mathbb{R}$ and $z_i \in [-L/2, L/2]$.
In our case, we choose $c=3$ with $z_1 = -L/2 , z_2 = 0, z_3 = L/2$, locating each Gaussian distribution at the center and the two extremes of the pore.
Thus, our external potential expression as in Eq.~\eqref{eq:rbf-potential} is fully parametrized as $\overline{V}(x|Q)$ where $Q = (Q^{1}_{1}, Q^{2}_{1}, Q^{1}_{2}, Q^{2}_{2}, Q^{1}_{3}, Q^{2}_{3})$.
In practice, if there are no prior assumptions on the smoothness of the external 
potential, standard cross-validation techniques can be used to arrive at the 
optimal number of basis functions.

We implement a Gaussian prior distribution, $P(Q) = \mathcal{N}(\overline{Q}, \Sigma_{Q})$ with mean $\overline{Q} = 0$ and diagonal covariance matrix $\Sigma_{Q}$.
We then select $\rho(x | Q)$ as the likelihood distribution considering its physical interpretation as the probability-density function,
\begin{align}
    P(\mathcal{D} | Q) = \prod_{i=1}^{M} \rho(x_i | Q). \label{eq:bayesian-likelihood}
\end{align}
The posterior distribution $P(Q|\mathcal{D})$ from Eq.~\eqref{eq:bayesian-inference} cannot be analytically computed.
To overcome this issue, we resort to an approximate method as explained below.

\subsection{Posterior sampling}

Obtaining samples from the posterior distribution $P(Q|\mathcal{D})$ might seem 
a challenging task when only information about the prior and the likelihood 
distributions is available.
Indeed, the product $P(Q)P(\mathcal{D}|Q)$ yields a non-normalized form of the 
posterior distribution as seen in Eq.~\eqref{eq:bayesian-inference}.
However, the posterior distribution can be approximated by considering samples 
from a Markov chain whose stationary distribution converges to the desired 
posterior representation.
Here, we apply the Metropolis-Hastings algorithm \cite{Hastings1970} as a Markov 
chain MC method to obtain a large number of samples whose distribution resembles the posterior distribution.

The Metropolis–Hastings algorithm considers local Markov
changes in the parameter space. 
The transition from $Q$ to $Q^{*}$ is governed by a probability distribution $P(Q^{*}|Q)$, which has to be defined.
By proposing $P(Q^{*}|Q)$, the current states of the Markov chain will evolve according to this transition probabiltiy.
Finally, the algorithm must guarantee that the resulting stationary distribution from the Markov chain produces the target density, the posterior distribution in this case.
To fulfill this condition, each transition is accepted with probability:
\begin{align}
    \alpha(Q^{*}|Q) = \min\left(1, \frac{P(Q^{*})P(\mathcal{D}|Q^{*})P(Q|Q^{*})}{P(Q)P(\mathcal{D}|Q)P(Q^{*}|Q)} \right). \label{eq:acceptance-probability}
\end{align}
Combining the proposed transitions $Q \rightarrow Q^{*}$ according to $P(Q^{*}|Q)$ and accepting them with probability $\alpha(Q^{*}|Q)$, the Metropolis-Hastings algorithm will generate a dependent chain $(Q_1, Q_2, \ldots, Q_{S})$ whose distribution approximates $P(Q|\mathcal{D})$.

As a proposal distribution $P(Q^{*}|Q)$, we consider a spherical-Gaussian distribution with zero mean and diagonal covariance matrix, $\mathcal{N}(0, \Sigma_{Q^{*}Q})$.
We assume the same step-transition level for each parameter of $Q$ leading to $\Sigma_{Q^{*}Q} = \delta \mathbf{I}$, where $\mathbf{I}$ is an identity matrix.
Note that the spherical-Gaussian distribution yields a symmetric transition probability $P(Q^{*}|Q) = P(Q|Q^{*})$.

Given that we have defined $P(\mathcal{D}|Q)$ as in Eq.~\eqref{eq:bayesian-likelihood}, $\rho(x|Q)$  must be computed for each sample $Q$ obtained from the Markov chain.
Therefore, after sampling from the proposal distribution $P(Q^{*}|Q)$ to obtain a new $Q$,
the external potential $\overline{V}(x | Q)$ is computed following Eq.~\eqref{eq:rbf-potential}.
Then, $\overline{V}(x | Q)$ is inserted in the minimization problem stated in Eq.~\eqref{eq:euler-lagrange} to recover $\rho(x|Q)$ in order to calculate the likelihood function, i.e., Eq.~\eqref{eq:bayesian-likelihood}.

\subsection{Results}

\begin{figure*}
\includegraphics[width=.9\textwidth]{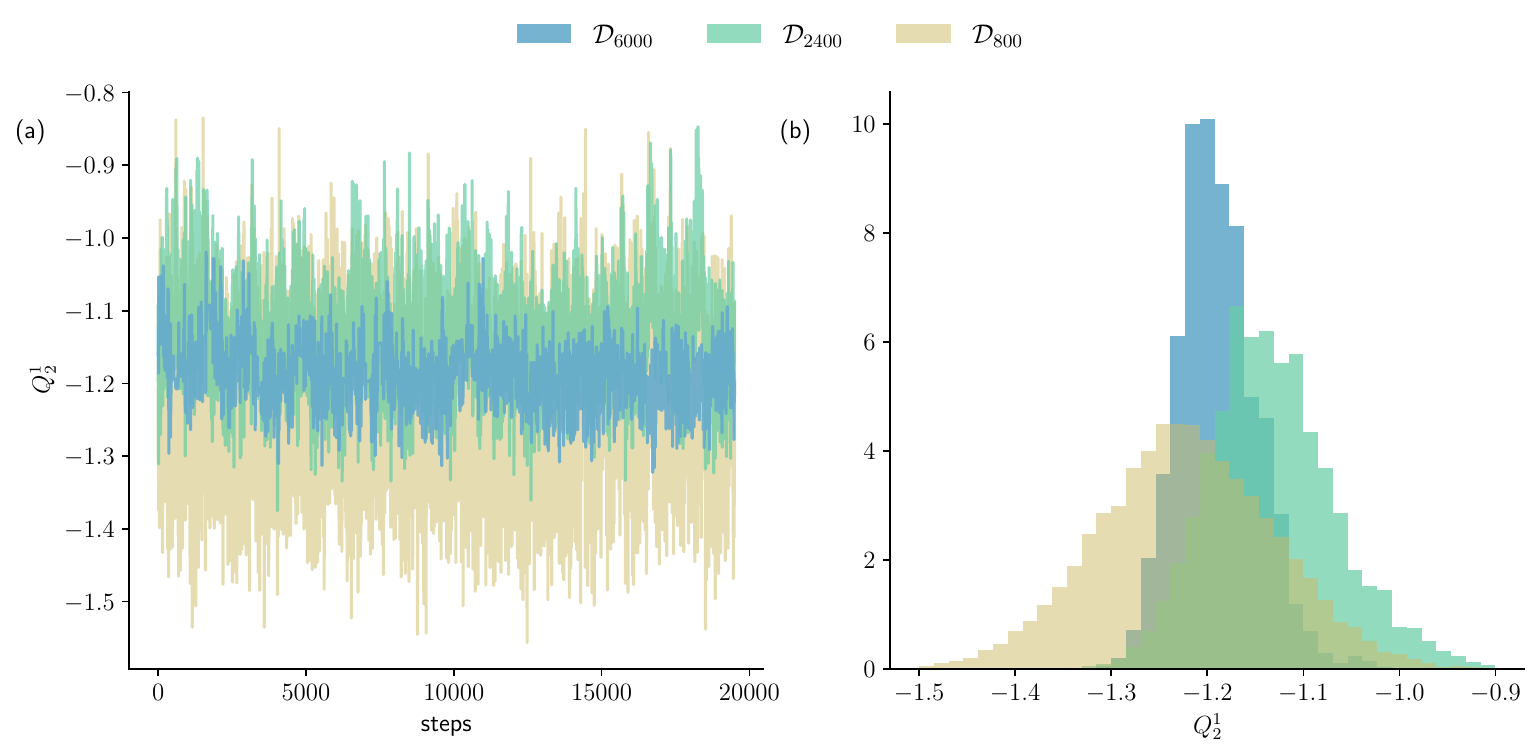}
\caption{\label{fig:chain_histogram_Q21} Illustration of posterior samples obtained from the Metropolis-Hastings algorithm for different input datatsets $\mathcal{D}_{800}, \mathcal{D}_{2400}$ and $\mathcal{D}_{6000}$. (a) Evolution of the Markov chains of $Q_2^{1}$ parameter for $2\times 10^{4}$ steps. The chains show rapidly variation and good mixing. (b) Histogram of the  $2\times 10^{4}$ posterior $Q_{1}^{2}$ samples. The posterior distribution contracts as more datapoints are included in the training dataset.}
\end{figure*}

To assess the external-potential inference capabilities of our
model we generate a dataset $\mathcal{D}$ with $N_{\text{conf}} = 3 \times 10^{5}$ and collect $800$ independent configurations to avoid particle 
configuration correlation, obtaining a training dataset $\mathcal{D}_{800}$ with $M \approx 5 \times 10^3$.
We then run the Metropolis–Hastings algorithm for $S= 10^{5}$ steps with $\delta=0.01$ in the transition distribution to validate the inference procedure.
The chain is then thinned every 5 steps to de-correlate it, obtaining $2\times 10^4$ final samples.
The resulting Markov chain of the $Q_{2}^{1}$ parameter for $\mathcal{D}_{800}$ is represented in Fig.~\ref{fig:chain_histogram_Q21}(a).
It is evident that the resulting chain yields a stationary distribution and a 
fast mixing.

\begin{figure*}
\includegraphics[width=1\textwidth]{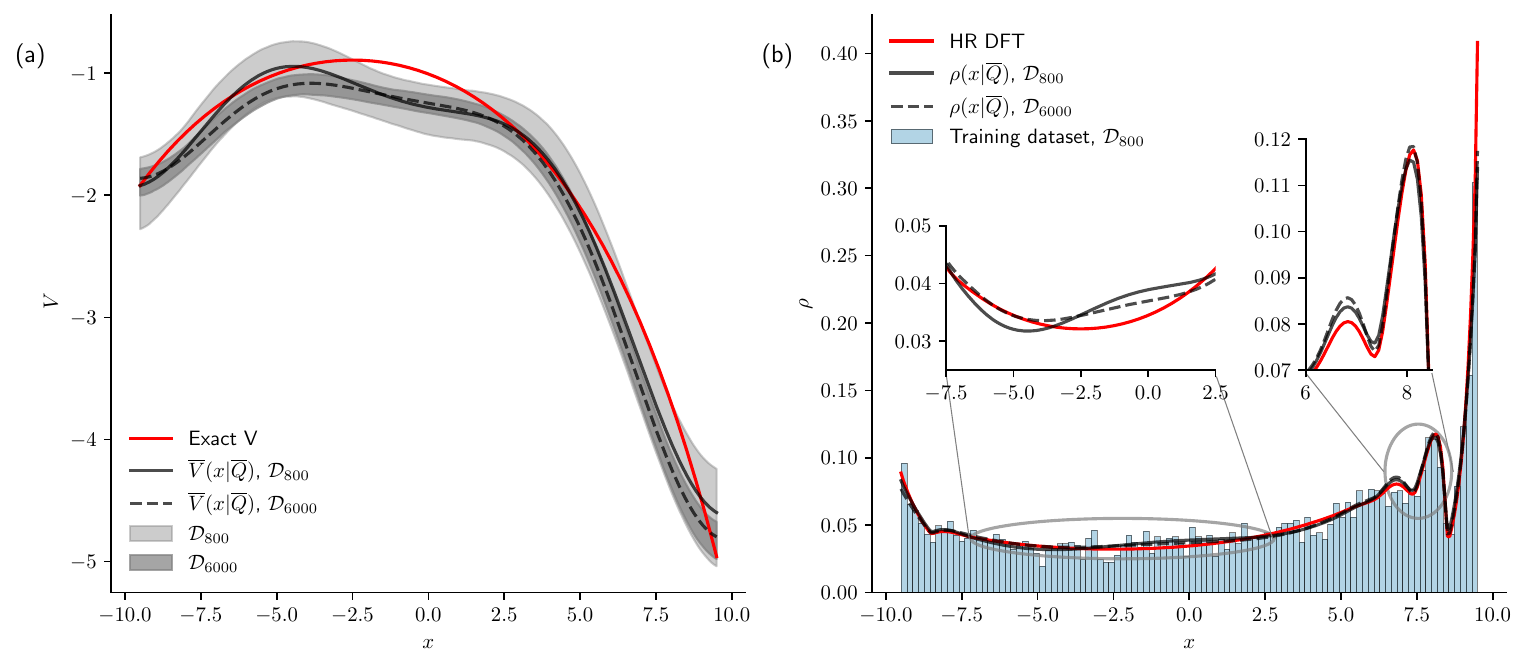}
\caption{\label{fig:potential_rho_d800} Illustration of the external potential inference with $\mathcal{D}_{800}$ and $\mathcal{D}_{6000}$. (a) Comparison between the predictive distribution $\overline{V}(x|Q)$ for training datasets $\mathcal{D}_{800}, \mathcal{D}_{6000}$ and exact external potential applied from Eq.~\eqref{eq:true-potential} [red line (online)]. The gray areas encloses the [1st, 99th] percentiles of the predictive distribution $\overline{V}(x|Q)$ for the stationary $Q$ samples obtained. The black solid line represents the predicted external potential for the mean parameters $\overline{Q} = (-4.55, \, 3.13, \, -1.22, \, 3.59,\, -1.85, \, 2.70)$ with $\mathcal{D}_{800}$, while the black dash line is the predicted external potential for the mean parameters $\overline{Q} = (-4.61, \, 3.11, \, -1.19, \,  4.01, \, -1.65, \, 2.81)$ with $\mathcal{D}_{6000}$. (b) Representation of the exact $\rho(x)$ obtained from analytical HR DFT [red line (online)], the predictive density profile distribution for the mean parameters $\overline{Q}$ (black solid line for $\mathcal{D}_{800}$ and black dash line for $\mathcal{D}_{6000}$) and the histogram of $\mathcal{D}_{800}$ used to infer the external potential.}
\end{figure*}

The predictive distribution $\overline{V}(x|Q)$ can be obtained by 
sampling for each $Q$.
The predictive distribution $\overline{V}(x|Q)$ and the exact external potential are superimposed in Fig.~\ref{fig:potential_rho_d800}(a), showing a good approximation of the external potential through the proposed Bayesian scheme.
It is worth mentioning that while the predicted external potential [shown as a 
black line in Fig.~\ref{fig:potential_rho_d800}(a)] 
$\overline{V}(x|\overline{Q})$ for the mean parameters $\overline{Q}$ 
in $\mathcal{D}_{800}$ does not fully match the exact potential
[shown as a red line (online) in Fig.~\ref{fig:potential_rho_d800}(a)], 
it yields a density profile $\rho(x|\overline{Q})$ for $\mathcal{D}_{800}$
almost identical to the exact $\rho(x)$, as demonstrated in Fig.~\ref{fig:potential_rho_d800}(b).
Recall that $\rho(x|\overline{Q})$ is computed solving the minimization problem in Eq.~\eqref{eq:euler-lagrange} inserting $\overline{V}(x|\overline{Q})$ as the external potential.

From Fig.~\ref{fig:potential_rho_d800}(a), the exact external potential 
[red line (online)] lies inside the $99\%$ probability region (light gray area) of the stationary predictive distribution $\overline{V}(x|Q)$ for 
$\mathcal{D}_{800}$ for the majority of the $x$-coordinates except for the small range $x \in [-2.5, 2]$.
In addition, the highest discrepancy between the exact potential 
[red line (online)] and the predicted external potential for the
mean parameters $\overline{V}(x|\overline{Q})$ for $\mathcal{D}_{800}$ 
(black solid line) occurs in this range.
Not surprisingly, this small range $[-2.5, 2]$ where $\overline{V}(x|Q)$ for 
$\mathcal{D}_{800}$ fails to represent the external potential corresponds
to the region with the lowest density, as shown in detail in 
Fig.~\ref{fig:potential_rho_d800}(b).
Nevertheless, this misrepresented region has
a negligible impact on the reconstruction of
$\rho(x)$ from the predictive distribution $\rho(x|Q)$, since $\rho(x)$ 
near the walls is inferred with high accuracy, capturing even the oscillatory 
behavior in $x \in (5.5, 9.5)$.

The high accuracy in the $\rho(x)$ approximation is achieved even with a 
training dataset $\mathcal{D}_{800}$ whose histogram [see 
Fig.~\ref{fig:potential_rho_d800}(b)] does not resemble the exact density profile as shown in Fig.~\ref{fig:exact_dft_simulation}.
This precision highlights the efficiency and robustness of our Bayesian
inference procedure. 
For the sake of evaluating the accuracy of our posterior distribution, we rerun 
the Metropolis–Hastings algorithm for two more training datasets, namely $\mathcal{D}_{2400}$ and $\mathcal{D}_{6000}$.
While $\mathcal{D}_{2400}$ provides $M \approx 1.5 \times 10^4$ particle coordinates, $\mathcal{D}_{6000}$ generates $M \approx 3.7 \times 10^4$ particle coordinates.
Evidently, the two datasets yield one order of magnitude above the baseline scenario, i.e., $\mathcal{D}_{800}$.
The resulting $Q_{2}^{1}$ chains and histograms for each dataset are represented in Fig.~\ref{fig:chain_histogram_Q21}.
The histograms in Fig.~\ref{fig:chain_histogram_Q21}(b) attest that increasing 
the number of particle coordinates reduces the variance of the estimated
$Q_{2}^{1}$ parameter obtaining contracted posterior distributions.
The same behavior is observed for the remaining $Q$ parameters.
This variance reduction propagates through the predictive distributions $\overline{V}(x|Q)$ and $\rho(x|Q)$, diminishing the inference uncertainty.
Such uncertainty reduction can be observed in 
Fig.~\ref{fig:potential_rho_d800}(a), where the $99\%$ probability region of 
$\overline{V}(x|Q)$ for $\mathcal{D}_{6000}$ (dark gray area) is narrower 
compared to the $99\%$ probability region of $\overline{V}(x|Q)$ for 
$\mathcal{D}_{800}$ (light gray area).

Furthermore, $\mathcal{D}_{6000}$ constitutes our most faithful scenario to
infer the external potential because we are considering more datapoints.
Indeed, the higher performance of the Bayesian inference
procedure for $\mathcal{D}_{6000}$ compared to $\mathcal{D}_{800}$ can be seen 
in the density comparison for $x \in [-7.5, 2.5]$ in 
Fig.~\ref{fig:potential_rho_d800}(b), where $\rho(x|\overline{Q})$ for 
$\mathcal{D}_{6000}$ (black dash line) approximates better the HR DFT density 
[red line (online)] compared to $\rho(x|\overline{Q})$ for $\mathcal{D}_{800}$ 
(black solid line).
On the other hand, the histogram obtained for $\mathcal{D}_{6000}$ in 
Fig.~\ref{fig:chain_histogram_Q21}(b) is within the estimated $Q_{2}^{1}$ range 
of the histogram generated by $\mathcal{D}_{800}$.
This suggests that, with fewer datapoints, we achieve a rough potential
distribution with higher standard deviation (uncertainty), although
the obtained distribution still represents a credible estimation of the
external potential.

We eventually determine a lower bound of particle coordinates
at which our Bayesian procedure becomes limited in its ability
to accurately infer the external potential. 
Considering a training dataset $\mathcal{D}_{360}$ with 
$M \approx 2.3 \times 10^3$ particle coordinates, we found that the 
resulting Markov chains of the $Q$ parameters quickly oscillate in the 
steady-state around value levels far 
from the ones obtained for $\mathcal{D}_{800}$.
This means that the $Q$ samples obtained from the stationary distribution for 
$\mathcal{D}_{360}$ are unable to efficiently recover the external potential.

\section{Conclusions}

The description of the microscopic interactions of a many-particle
system is a long-standing issue across several scientific and
engineering disciplines. 
The presence of an external potential in a many-particle system dramatically 
alters these interactions. 
For example, the governing external field plays a crucial role in fluid 
substrate interactions like adsorption or wetting transitions. 
Therefore, knowing the external potential exerted over a many-particle
system is key to making advances in the theoretical-computational
exploration of novel complex materials.

Our overarching objective here is to push forward existing techniques
to uncover the potential function between the external field
and the many-particle ensemble. 
For this purpose we adopted a Bayesian framework, exemplified with a 
grand-canonical 1D HR system in a confined geometry under the influence of an 
external potential. 
The training dataset to infer the external potential 
is generated using a grand-canonical MC simulation. 
From the simulated data, the Bayesian framework reconstructs the external
potential, which is embedded into the classical DFT formulation
to generate the equilibrium-density function. 
The resulting density is then compared with the exact density output produced by
the DFT apparatus using the true external potential term. 
Hence, the validity of our framework is contingent upon the degree to
which the inferred potential accurately leads to the same density
function. 
Despite the simplicity of our prototype, its core functionality
resembles many real-world applications involving adsorption
or capillarity features, and our statistical learning framework
can be employed for empirical-microscopic observations of systems
exhibiting a similar configuration to our computational set-up,
thus, driving a more efficient, rational, and systematic study of
fluid-substrate interactions.

The primary benefit of our Bayesian methodology lies in its
ability to quantitatively assess uncertainty throughout all modeling
hierarchy levels. 
This uncertainty quantification is rooted in the potential-parameter estimation 
as well as in the density-profile distribution obtained after inserting the 
inferred potential in the classical DFT formulation. 
In fact, we have demonstrated the existing trade-off between the degree of 
uncertainty in the modeling and the amount of available data. 
This uncertainty-data-availability relationship and the probabilistic 
description of the results motivate a much improved interpretation of the 
modeling output over classical ML techniques such as neural networks.

At the heart of our framework is the functional estimation of
the external potential acting on the system at hand for which a basic
Gaussian RBF has been proposed. 
Here, only three terms in the RBF, six parameters in total, are implemented to 
approximate the external potential. 
This small number of parameters is a direct consequence of the simple 
external-potential functional form we have imposed.
Further improvements of the external potential evaluation can be
carried out considering more terms in the RBF expansions or a polynomial
expansion providing capabilities to represent more complex
potential functions.

While our learning framework has shown a very good performance
in reconstructing the external potential, there is scope
for improvement and further refinement. 
First, of particular interest would be extension to higher-dimensional 
many-particle systems, which, at least theoretically, seems straightforward to 
implement if the one-body density is defined appropriately. 
However, a higher dimensional configuration would incur greater computational
costs, thus hindering the sampling procedure. 
This computational constraint should be painstakingly addressed. 
Another interesting extension would be to utilize appropriate coarse-graining
techniques to generalize the interparticle-interaction potential to
account for subtle details of complex microscopic interactions. 
Such techniques typically decompose the external potential function into
different terms serving specific purposes. 
Future research could be devoted to inferring each component of the external 
potential decomposition using the Bayesian inference framework developed
here.

\section*{ACKNOWLEDGMENTS}

A.M. was supported by Imperial College London President’s
Ph.D. Scholarship scheme. P.Y. was supported by Wave 1 of
The UKRI Strategic Priorities Fund under EPSRC Grant No.
EP/T001569/1, particularly the “Digital Twins for Complex Engineering
Systems” theme within that grant, and The Alan Turing Institute. 
S.K. was supported by ERC through Advanced Grant
No. 247031 and EPSRC through Grant Nos. EP/L025159 and
EP/L020564.

\bibliography{references}

\end{document}